\documentclass[11pt, a4paper]{article}
\usepackage{jcappub}
\usepackage{bm}
\usepackage{graphicx}
\usepackage{epstopdf}

\title{Natural Warm Inflation}

\author{Luca Visinelli}
\affiliation{Department of Physics and Astronomy, University of Utah, 115 South 1400 East \#201, Salt Lake City, Utah 84112-0830, USA}
\emailAdd{u0583682@utah.edu}

\abstract{We derive the requirements that a generic axion-like field has to satisfy in order to play the role of the inflaton field in the warm inflation scenario. Compared to the parameter space in ordinary Natural Inflation models, we find that the parameter space in our model is enlarged. In particular, we avoid the problem of having an axion decay constant $f$ that relates to the Planck scale, which is instead present in the ordinary Natural Inflation models; in fact, our model can easily accommodate values of the axion decay constant that lie well below the Planck scale.}

\arxivnumber{1107.3523}

\begin{document}
\maketitle

\section{Introduction}

Inflation \cite{kazanas, starobinsky, guth, sato, albrecht, linde} provides a mechanism for generating the inhomogeneities observed in the Cosmic Microwave Background Radiation (CMBR) \cite{mukhanov, guth_pi, hawking, starobinsky1, bardeen1}, and explains the observed flatness, homogeneity, and the lack of relic monopoles that posed severe problems in the standard Big-Bang cosmology \cite{linde_book, kolb_book}. Realistic microphysical models of inflation, in which the expansion of the universe is governed by the energy density of the inflaton field $\phi$, are complicated by the requirement  that the inflaton potential $U(\phi)$ be very flat in order to explain the anisotropies observed in the CMBR \cite{limits_self_coupling}.

Natural Inflation \cite{natural_inflation, natural_inflation1, natural_inflation2} is a viable model in which the inflaton is identified with an axion-like particle: in fact, the shift symmetry $\phi \to \phi$ + const. present in axionic theories assures a flat inflaton potential. Although Natural Inflation is well-motivated and it is consistent with the WMAP measurements \cite{natural_inflation2}, it is not an easy task to embed this model in fundamental theories like string theory \cite{banks}, the main complication coming from the fact that the energy scale $f$ at which the shift symmetry spontaneously breaks must be $f>0.6~M_{\rm Pl}$, in order to agree with the constraints on the scalar spectral index $n_s$.

In this paper we show that the energy scale $f$ can be as low as the Grand Unification Theory (GUT) scale $\Lambda_{\rm GUT}~\sim~10^{16}{\rm~GeV}$ if Natural Inflation is considered in the context of Warm Inflation (\cite{berera}; see also Refs.~\cite{hosoya, moss_ewi, lonsdale, yokoyama, liddle_ewi}); we refer to this as the Natural Warm Inflation (NWI) model.

Since axion-like particles arise in generic four-dimensional models \cite{masso, coriano}, string theory compactifications \cite{axion_strings}, and generic Kaluza-Klein theories \cite{axion_KK}, and possess attractive features for inflation models like a flat potential already embedded in the theory, such particles have been extensively discussed in the inflation literature \cite{axion_like, mohanty, kim}. In particular, an early attempt at lowering the Natural Inflation scale $f$ in the Warm Inflation scenario has been discussed in Ref.~\cite{mohanty}. However, the authors in Ref.~\cite{mohanty} treat the temperature $T$ of the primordial plasma as an independent variable whereas, as remarked below in this paper, the temperature $T$ is not an independent quantity once the energy density of the relativistic species $\rho_r$ is specified \cite{berera}.

This paper is organized as follows. We first fix our notation for the Warm Inflation scenario and for the axion particle physics in Sections~\ref{The warm inflation scenario} and~\ref{Axion-like particles}, respectively. In Section~\ref{Warm natural inflation} we analyze the dynamic of the inflaton field using the slow-roll conditions and number of e-folds for sufficient inflation, and we study the parameter space of our NWI model. In Section~\ref{Perturbations from inflation} we discuss the bounds on the NWI model resulting from the measurements of cosmological parameters, in the light of the Wilkinson Microwave Anisotropy Probe (WMAP) mission plus baryon acoustic oscillations (BAO) and supernovae (SN) data \cite{komatsu}. Finally, discussions and conclusions are drawn in Section~\ref{discussion}.

\section{The warm inflation scenario}\label{The warm inflation scenario}

In the warm inflation scenario the inflaton field appreciably converts into relativistic matter (from here on referred to as ``radiation'') during the inflationary period. This mechanism is parametrized by the appearance of a dissipative term $\Gamma$ in the dynamics of the inflaton field. In the following we reasonably assume \cite{berera} that radiation thermalizes on a time scale much shorter than $1/\Gamma$. The energy density in radiation is thus
\begin{equation}\label{definition_radiation}
\rho_r = \frac{\pi^2}{30}\,g_*(T)\,T^4,
\end{equation}
where $g_*(T)$ is the number of relativistic degrees of freedom of radiation at temperature $T$. Here we do not specify $g_*(T)$ in the equations, but in the figures we will always use $g_*(T) = 228.75$, corresponding to the number of relativistic degrees of freedom in the minimal supersymmetric Standard Model.

Warm inflation is achieved when thermal fluctuations dominate over quantum fluctuations, or \cite{berera}
\begin{equation} \label{condition_warm_inflation}
H(T) < T,
\end{equation}
where $H$ is the Hubble expansion rate at temperature $T$. The effectiveness at which the inflaton converts into radiation is measured by the ratio
\begin{equation} \label{effectiveness}
Q = \frac{\Gamma}{3H};
\end{equation}
for $Q \gg 1$ a strongly dissipative regime is achieved, while $Q <1$ represents the weak regime of warm inflation. Throughout this Section we will present the general equations for warm inflation, focusing on the case $Q \gg 1$ in the subsequent Sections when specified.

In the following, we model the inflaton field with a scalar field $\phi = \phi(x)$ minimally coupled to the curvature and moving in a potential $U = U(\phi)$. The evolution of the inflaton field in a Friedmann-Robertson-Walker metric is described by
\begin{equation} \label{eq_motion}
\ddot{\phi} + (3H+\Gamma)\dot{\phi} + U_\phi = 0,
\end{equation}
where a dot indicates the derivation with respect to the cosmic time $t$ and $U_\phi = \partial U/\partial\phi$. The conservation of the total energy of the system imposes that the radiation energy density $\rho_r$ satisfies
\begin{equation} \label{energy_conservation_radiation}
\dot{\rho_r} + 4H\rho_r = \Gamma\,\dot{\phi}^2,
\end{equation}
with the term on the RHS of Eq.~(\ref{energy_conservation_radiation}) describing the effectiveness of conversion of the inflaton field into radiation.

The total energy density of the system at any time is $\rho_{\rm tot} = \dot{\phi}^2/2 + U(\phi, T)$, where $U(\phi,T)$ is an effective potential that accounts for temperature effects. As discussed in Ref.~\cite{moss}, a requirement for building a consistent model of warm inflation is that finite temperature effects on the inflaton potential must be suppressed. This requirement makes possible to separate the effective potential of the inflaton as $U(\phi, T) = U(\phi) + \rho_r(T) = U + \rho_r$. We thus have
\begin{equation}
\rho_{\rm tot} = \frac{1}{2}\dot{\phi}^2 + U + \rho_r,
\end{equation}
and the corresponding Friedmann equation reads
\begin{equation} \label{friedmann}
H^2 = \frac{8\pi}{3M_{\rm Pl}^2}\left(\frac{1}{2}\dot{\phi}^2+U+\rho_r\right),
\end{equation}
where $M_{\rm Pl} = 1.221\times 10^{19}{\rm~GeV}$ is the Planck mass.

Inflation takes place when the potential $U$ is approximately flat and the potential energy dominates over all other forms of energy, so that the Hubble expansion rate $H$ is constant. During this period, which is known as the slow-roll regime of the inflaton field, higher derivatives in Eqs.~(\ref{eq_motion}) and~(\ref{energy_conservation_radiation}) can be neglected. In formulas,
\begin{equation}\label{slow_roll_conditions}
\ddot{\phi} \ll H \,\dot{\phi},\quad\hbox{and}\quad \dot{\rho}_r \ll H\,\rho_r.
\end{equation}
In this regime Eq.~(\ref{eq_motion}) reads
\begin{equation} \label{eq_motion_slow_roll}
\dot{\phi} \simeq -\frac{U_\phi}{3H+\Gamma},
\end{equation}
where here and in the following we use the symbol ``$\simeq$'' for an equality that holds only in the slow-roll regime. Eqs.~(\ref{energy_conservation_radiation}) and~(\ref{friedmann}) in the slow-roll regime respectively read
\begin{equation} \label{energy_conservation_sl}
\rho_r \simeq \frac{3Q}{4}\,\dot{\phi}^2,
\end{equation}
and
\begin{equation} \label{friedmann_sl}
H^2 \simeq \frac{8\pi G}{3}\,U,
\end{equation}
so that a shallow potential $U$ gives rise to a nearly constant expansion rate $H$.

The slow-roll regime can be parametrized by a set of slow-roll parameters $\epsilon$, $\eta$ and $\beta$, defined by
\begin{equation}\label{slow_roll_parameters}
\epsilon = \frac{1}{16\pi G}\left(\frac{U_\phi}{U}\right)^2, \quad \eta = \frac{1}{8\pi G}\,\frac{U_{\phi\phi}}{U},\quad\beta = \frac{1}{8\pi G}\left(\frac{\Gamma_{\phi}\,U_\phi}{\Gamma\,U}\right).
\end{equation}
To assure the conditions expressed in Eq.~(\ref{slow_roll_conditions}), we first differentiate Eqs.~(\ref{eq_motion_slow_roll}),~(\ref{energy_conservation_sl}) and~(\ref{friedmann_sl}), obtaining
\begin{equation}
\frac{\dot{H}}{H^2} \simeq -\frac{\epsilon}{1+Q},
\end{equation}
\begin{equation}
\frac{\ddot{\phi}}{H\,\dot{\phi}} \simeq -\frac{1}{1+Q}\left(\eta-\beta+\frac{\beta-\epsilon}{1+Q}\right),
\end{equation}
\begin{equation}
\frac{\dot{\rho}_r}{H\,\rho_r} \simeq -\frac{1}{1+Q}\left(2\eta-\beta-\epsilon+2\frac{\beta-\epsilon}{1+Q}\right).
\end{equation}
Since the right hand side in the last three expressions gives the size of the terms neglected in the slow-roll approximation, we meet the conditions in Eq.~(\ref{slow_roll_conditions}) by demanding that these terms be small (see also Refs.~\cite{taylor_berera, hall, moss}),
\begin{equation} \label{slow_roll}
\epsilon \ll 1+Q,\quad |\eta| \ll 1+Q,\quad |\beta| \ll 1+Q.
\end{equation}
Eq.~(\ref{slow_roll}) is a generalization of the slow-roll conditions in the cool inflation that takes into account the parameter $Q$; when $Q \ll 1$ the dissipation term can be neglected and the slow-roll conditions reduce to the usual requirements in the cool inflation.

In the warm inflation scenario, we can constraint the scale of the potential $U$ by combining the definition of $\rho_r$ in Eq.~(\ref{definition_radiation}) with the requirements that $U \gg \rho_r$ and Eq.~(\ref{condition_warm_inflation}) to obtain the constraint
\begin{equation}
U^{1/4} \ll \left(\frac{135}{32\,g_*(T)\,\pi^4}\right)^{1/4}\,M_{\rm Pl} = 5.57\,g_*^{-1/4}(T) \times 10^{18}{\rm~GeV}.
\end{equation}
This is not a stringent bound, since in most theories of grand unification the inflaton potential is related to the unification scale, $U^{1/4} \sim 10^{16}$ GeV.

\section{Axion-like particles}\label{Axion-like particles}

The axion \cite{PQ, weinberg, wilczek} is the pseudo Nambu-Goldstone boson associated with the breaking of the Peccei-Quinn symmetry proposed as a solution to the strong CP problem. The theory of the QCD axion predicts that the axion mass at zero temperature $m_a$ is inversely proportional to the axion decay constant $f_a$, with the constant of proportionality related to the QCD scale $ \Lambda_{\rm QCD}$. In formulas,
\begin{equation}\label{relation_qcd}
\Lambda_{\rm QCD} \sim \sqrt{m_a\,f_a}.
\end{equation}
A generic axion-like theory is a generalization of the original axion theory, in which the mass $m_\phi$ and the decay constant $f$ of the axion theory are not related as in Eq.~(\ref{relation_qcd}) but rather by a more generic relation, 
\begin{equation}\label{def_lambda}
\Lambda = \sqrt{m_{\phi}\,f},
\end{equation}
where the scale $\Lambda$ is fixed by the underlying theory. In fact, axion-like particles arise naturally whenever an approximate global symmetry is spontaneously broken, like in generic four-dimensional models \cite{masso, coriano}, string theory compactifications \cite{axion_strings} and generic Kaluza-Klein theories \cite{axion_KK}; all of these theories predict specific values for $\Lambda$. Axion-like particles mainly differ from the original QCD axion because the energy scale $\Lambda$ differs from $\Lambda_{\rm QCD}$: depending on the underlying theory, the scale $\Lambda$ assumes values that range up to the GUT scale $\Lambda_{\rm GUT}~\sim~10^{16}$ GeV. Here the scale $\Lambda$ is not fixed, so that the parameters $m_\phi$ and $f$ are unrelated \cite{masso1}, and Eq.~(\ref{def_lambda}) merely defines $\Lambda$ once both $m_\phi$ and $f$ are given.

It is well known that axion-like particles serve as suitable candidates for the inflaton field because the axion potential
\begin{equation} \label{potential}
U = U(\phi) = \Lambda^4 \left[1 + \cos\left(N\,\frac{\phi}{f}\right)\right],
\end{equation}
with $N$ integer, is naturally flat whenever $\phi/f$ is not close to $\pi/N$. Moreover, in a generic inflaton model it is required that the coupling term $\lambda_\phi$ multiplying the quartic self-interaction $\phi^4$ in the inflaton lagrangian satisfies $\lambda_\phi \lesssim 10^{-}$ \cite{limits_self_coupling}. From Eq.~(\ref{potential}), the quartic self-interaction term in the NWI model is
\begin{equation}\label{self_interaction}
\lambda_\phi = \frac{m_\phi^2}{24f^2}.
\end{equation}
In Section~\ref{Scalar power spectrum} we will prove that the self-interaction term in Eq.~(\ref{self_interaction}) fulfills the requirement from observations, see Figure~\ref{plotLambda}.

{}Axion-like theories provide a model for the decay rate of $\phi$. In fact, an axion-like particle $\phi$ couples to a gauge field $F$ via the Lagrangian term
\begin{equation}
\mathcal{L}_{\phi F} = \frac{g^2}{32\pi^2}\,\phi\,F\,\tilde{F},
\end{equation}
where $\tilde{F}$ is the dual of the gauge field $F$ and  $g$ is a coupling constant. For example, in the invisible axion theory $F$ can be either the gluon or the electromagnetic field. In the latter case, axion-like particles decay in a pair of photons with a rate (see Ref.~\cite{raffelt})
\begin{equation}\label{decay_constant}
\Gamma =g^2\,\frac{m_\phi^3}{f^2}.
\end{equation}

Due to instanton effects, axion-like particles have a temperature dependent mass as \cite{gross}
\begin{equation} \label{axion_mass_T}
m_\phi(T) = m_\phi\,\begin{cases}
1 & \hbox{$T < \Lambda$},\\
\left(\frac{\Lambda}{T}\right)^4 & \hbox{$T>\Lambda$}.
\end{cases}
\end{equation}
In principle, this leads to a non-zero value of $\Gamma_\phi = d\Gamma/d\phi$ when $T>\Lambda$. However, as we will show in Section~\ref{parameter_space}, the NWI model is consistent only with the case $T < \Lambda$ where the axion-like mass is constant. Thus, temperature effects on the axion-like mass can be neglected; in the following we set $\Gamma_\phi = 0$ and, consequently, $\beta = 0$.

\section{Warm natural inflation} \label{Warm natural inflation}

From now on we identify the axion-like particle $\phi$ with the inflaton, using the flat potential in Eq.~(\ref{potential}) to describe the dynamics for the inflaton field in the early universe. Here, we will take $N = 1$ in Eq.~(\ref{potential}), so that the potential has a unique minimum at $\phi = \pi f$; the inflaton field relaxes towards this minimum in its evolution. In the following, we indicate the inflaton with $\phi$, bearing in mind that we have assumed that $\phi$ is also a pseudo-scalar particle for which the formulas in Section~\ref{Axion-like particles} apply.

Furthermore, from here on we are interested only in the strongly dissipative regime of warm inflation, $Q \gg 1$. In this limit, the only parameters in the theory are the inflaton mass $m_\phi$, the decay constant $f$ and the dissipation term $\Gamma$. In table 1 we have collected the most important quantities of the theory in terms of these parameters

\begin{table}[h!]
\begin{tabular}{ll}
{\bf Quantity} \hspace{4em} & {\bf Equation(s) used} \hspace{12em}\\
\hline\\
$U(\phi) = m_\phi^2\,f^2\,(1+\cos\phi/f),$ & Eq.~(\ref{potential});\\
$U_\phi(\phi) = -m_\phi^2\,f\,\sin\phi/f,$ & derived from Eq.~(\ref{potential});\\
$H \simeq \frac{m_\phi\,f}{M_{\rm Pl}}\,\sqrt{\frac{8\pi}{3}\,(1+\cos\phi/f)},$ & Eqs.~(\ref{friedmann_sl}) and~(\ref{potential});\\
$\dot{\phi} \simeq \frac{m_\phi^2\,f}{\Gamma}\,\sin\phi/f,$ & Eqs.~(\ref{eq_motion_slow_roll}),~(\ref{friedmann_sl}) and~(\ref{potential});\\
$Q \simeq \frac{\Gamma\,M_{\rm Pl}}{m_\phi\,f}\,\sqrt{\frac{1}{24\pi(1+\cos\phi/f)}},$ & Eqs.~(\ref{potential}) and~(\ref{effectiveness_sl});\\
$\rho_r \simeq \sqrt{\frac{3}{128\pi}}\frac{m_\phi^3\,f\,M_{\rm Pl}}{\Gamma}\,\frac{\sin^2\phi/f}{\sqrt{1+\cos\phi/f}},$ & Eqs.~(\ref{eq_motion_slow_roll}),~(\ref{energy_conservation_sl}),~(\ref{friedmann_sl}) and~(\ref{potential});\\
$T^4 \simeq \sqrt{\frac{675}{32\pi^5}}\frac{m_\phi^3\,f\,M_{\rm Pl}}{\Gamma\,g_*(T)}\,\frac{\sin^2\phi/f}{\sqrt{1+\cos\phi/f}},$ & Eqs.~(\ref{definition_radiation}),~(\ref{eq_motion_slow_roll}),~(\ref{energy_conservation_sl}),~(\ref{friedmann_sl}) and~(\ref{potential});\\
$\epsilon = \frac{1}{16\pi\,G\,f^2}\,\frac{\sin^2\phi/f}{(1+\cos\phi/f)^2},$ & Eqs.~(\ref{slow_roll_parameters}) and~(\ref{potential});\\
$\eta \simeq  -\frac{1}{8\pi\,G\,f^2}\,\frac{\cos\phi/f}{1+\cos\phi/f},$ & Eqs.~(\ref{slow_roll_parameters}) and~(\ref{potential}).\\
\\
\hline\\
\end{tabular}
\label{table1}
\caption{Expressions for some derived quantities in the theory, valid during slow-roll and $Q \gg1$.}
\end{table}
We now examine the constraints on the NWI model coming from the slow-roll conditions in Eq.~(\ref{slow_roll}), the requirement for sufficient inflation, and the WMAP measurements on the power spectrum of density and tensor perturbations.

\subsection{Slow-roll conditions}

During warm inflation the values of the slow-roll parameters $\epsilon$ and $\eta$ must both be smaller than $Q$, see Eq.~(\ref{slow_roll}) with $Q \gg 1$. Here we neglect the condition $\beta \ll Q$ because in this model $\beta = 0$. Inflation ends when one of these two slow-roll conditions is violated.

Writing the parameter $Q$ during slow-roll as
\begin{equation} \label{effectiveness_sl}
Q \simeq \frac{\Gamma}{\sqrt{24\pi G\,U}},
\end{equation}
the slow-roll conditions $\epsilon \ll Q$ and $\eta \ll Q$ read
\begin{equation}
\epsilon = \frac{1}{16\pi G}\left(\frac{U_\phi}{U}\right)^2 \ll \frac{\Gamma}{\sqrt{24\pi G\,U}},
\end{equation}
and
\begin{equation}
\eta = \frac{U_{\phi\phi}}{8\pi \,G \,U} \ll \frac{\Gamma}{\sqrt{24\pi G\,U}},
\end{equation}
Using the expression for the potential $U(\phi)$ in Eq.~(\ref{potential}), these slow-roll conditions give respectively
\begin{equation} \label{slow_roll_epsilon1}
\frac{1-\cos\phi/f}{\sqrt{1 + \cos\phi/f}} \ll \alpha,
\end{equation}
and
\begin{equation} \label{slow_roll_eta1}
\frac{-\cos\phi/f}{\sqrt{1 + \cos\phi/f}} \ll \frac{\alpha}{2},
\end{equation}
where we have defined the combination
\begin{equation} \label{def_alpha}
\alpha \equiv \sqrt{\frac{32\pi}{3}}\frac{\Gamma\,f}{M_{\rm Pl}\,m_{\phi}} = 4.74\,\Gamma_{12}\,f_{16}\,m_{\phi \,9}^{-1}.
\end{equation}
Here and in the following, a number $y$ indexing some quantity with units of energy indicates that such quantity has been divided by $10^y$ GeV: for example, $f_{16} = f/10^{16}$GeV.

The slow-roll regime ends when the field $\phi$ reaches a value $\phi_f$ for which one of the conditions $\epsilon \ll Q$ or $\eta \ll Q$ is no longer satisfied. We have checked numerically that in general the second condition $\eta \ll Q$ is tighter than the first one $\epsilon \ll Q$, but for $\alpha \gg 1$ the two conditions give the same result,
\begin{equation}  \label{definition_phi2}
\phi_f= f\,\left(\pi - \frac{\sqrt{8}}{\alpha}\right).
\end{equation}
Since observations favor a large value of $\alpha > 10$ (see below), in this paper we use Eq.~(\ref{definition_phi2}) to define $\phi_f$.

\subsection{Number of E-folds}

The number of e-folds is defined as
\begin{equation} \label{number_efoldings}
N_e \equiv \ln(a_2/a_1) = \int_{t_1}^{t_2} H dt,
\end{equation}
where $a_1$ and $a_2$ are the values of the scale factor $a(t)$ appearing in the Friedmann metric when inflation begins and ends, respectively. Sufficient inflation requires
\begin{equation}
N_e > 60.
\end{equation}
During the inflationary stage, the value of the inflaton field decreases from the initial value $\phi_i$ to the value at the end of inflation $\phi_f$ defined via Eq.~(\ref{definition_phi2}). We now derive a relation between $\phi_i$ and $\phi_f$, using the definition of $N_e$ above. In the case of a slow-rolling of the inflaton and in the strongly dissipative regime, Eq.~(\ref{number_efoldings}) reads
\begin{equation} \label{number_efoldings1}
N_e \simeq -\int_{\phi_i}^{\phi_f}\frac{H\,\Gamma}{U_\phi}d\phi,
\end{equation}
where we used Eq.~(\ref{eq_motion_slow_roll}) with $\Gamma \gg 3H$. Using the axion-like potential in Eq.~(\ref{potential}) and the Friedmann equation we obtain
\begin{equation}
N_e = \sqrt{\frac{8\pi}{3}}\,\frac{\Gamma\,f}{m_\phi\,M_{\rm Pl}}\,\int_{\phi_i/f}^{\phi_f/f}dx\,\frac{\sqrt{1+\cos x}}{\sin x} = \frac{\alpha}{\sqrt{2}}\,\ln\,\tan \frac{x}{4}\,\bigg|_{\phi_i/f}^{\phi_f/f},
\end{equation}
or
\begin{equation} \label{relation_phi1_phi2}
\tan\frac{\phi_f}{4f}= \tan\frac{\phi_i}{4f}\,{\rm Exp}\left[\frac{\sqrt{2}\,N_e}{\alpha}\right].
\end{equation}
In general, we use Eqs.~(\ref{definition_phi2}) and~(\ref{relation_phi1_phi2}) to obtain the value of $\phi_i/f$, 
\begin{equation}\label{definition_phi1}
\frac{\phi_i}{f} = 4\arctan \left[\frac{1-\tan\frac{1}{\sqrt{2}\alpha}}{1+\tan\frac{1}{\sqrt{2}\alpha}}\,{\rm Exp}\left(-\frac{\sqrt{2}N_e}{\alpha}\right)\right]
\end{equation}
The appearance of trigonometric functions is due to the shape of the potential $U(\phi)$, that differs from a pure quadratic one and contains a cosine function itself.

\begin{figure}[h!]
\begin{center}
\includegraphics[width=10cm]{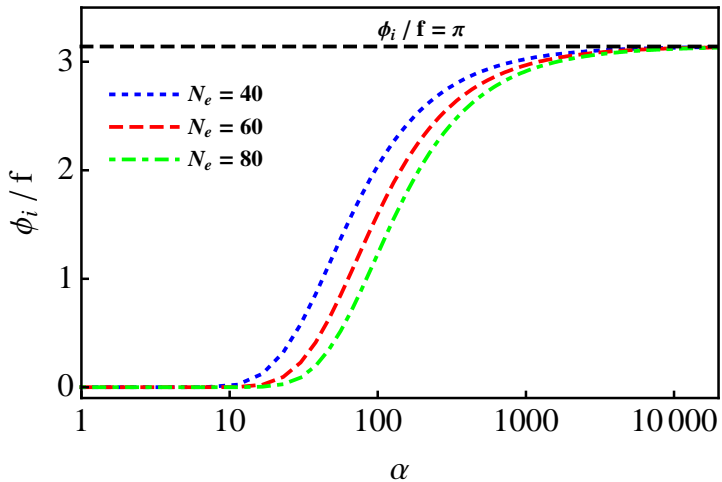}
\caption{The initial value of the angle $\phi_i/f$ as a function of $\alpha = \sqrt{32\pi/3} \,\Gamma f/m_\phi M_{\rm Pl}$ for different value of the numbers of e-folds $N_e$. Blue dotted: $N_e=40$; Red dashed: $N_e=60$; Green dot-dashed: $N_e=80$. Also shown is the line $\phi_i = \pi f$ (Black dashed line).}
\label{phi1_f}
\end{center}
\end{figure}

In Figure~\ref{phi1_f} we show the value of $\phi_i/f$ given in  Eq.~(\ref{definition_phi1}) as a function of $\alpha$. Considering the three parameters $m_\phi$, $f$ and $\Gamma$ from which $\alpha$ depends, moving towards greater values of $\alpha$ corresponds to a decrease in $m_\phi$ or to an increase in $f$ or $\Gamma$, once the other two parameters have been fixed.

\begin{figure}[h!]
\begin{center}
\includegraphics[width=10cm]{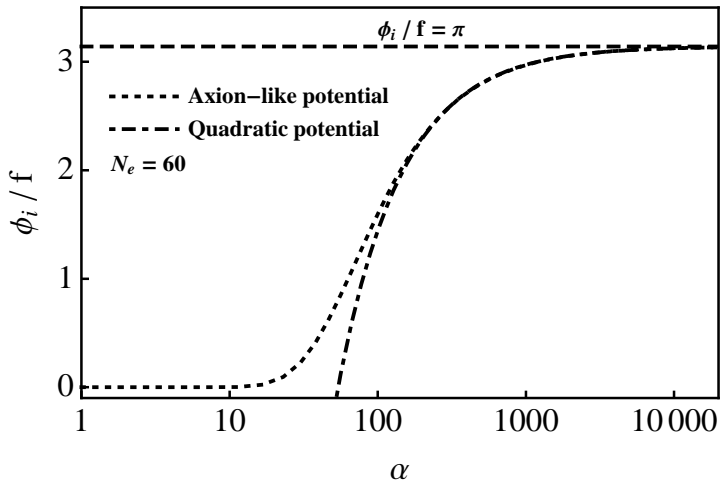}
\caption{The initial value of the angle $\phi_i/f$ as a function of $\alpha = \sqrt{32\pi/3} \,\Gamma f/m_\phi M_{\rm Pl}$ for the case of the axion-like potential in Eq.~(\ref{potential}) (black dotted line) and for the quadratic potential in Eq.~(\ref{quadratic_potential}) that approximates Eq.~(\ref{potential}) around $\phi_i \sim \pi\,f$ (black dot-dashed line). We fixed $N_e = 60$.}
\label{phi1_f2}
\end{center}
\end{figure}

For larger values of $\alpha$, the initial angle $\phi_i/f$ approaches $\pi$ and the axion-like potential is not distinguishable from a pure quadratic one. This fact is shown in Figure~\ref{phi1_f2}, where we compare the value of the angle $\phi_i/f$ with $N_e = 60$, together with the value of $\phi_i$ if the inflaton potential were a pure quadratic one,
\begin{equation} \label{quadratic_potential}
U_{\rm quad}(\phi) = \frac{\Lambda^4}{2}\,\left(\pi-\frac{\phi}{f}\right)^2,
\end{equation}
instead of Eq.~(\ref{potential}).
For $\alpha \gtrsim N_e$, it is
\begin{equation} \label{phi1_approx}
\phi_i \approx \pi - N_e \sqrt{8}/\alpha,
\end{equation}
and the dynamics of the field can no longer discern between the two potentials $U(\phi)$ and $U_{\rm quad}(\phi)$.

\subsection{Parameter space of the NWI} \label{parameter_space}

We have derived our results in the strongly dissipating regime $\Gamma \gg 3H$: this constraints the parameters as
\begin{equation} \label{constraint_G}
\alpha \gg 16\pi\,\left(\frac{f}{M_{\rm Pl}}\right)^2\,\sqrt{1+\cos\phi_i}.
\end{equation}
This constraint is not severe for values of $f$ around the GUT scale, which is the region of interest in this paper. In fact, for $f\sim 10^{16}$ GeV a numerical solution of Eq.~(\ref{constraint_G}) is
\begin{equation}\label{cnstr1}
\alpha \gg 4.8\times 10^{-5}\,f_{16}^2,
\end{equation}
independently of $N_e$. This bound is much more loose than the requirement of flatness of $U(\phi)$, which yielded $\alpha > 1$. When $f \sim M_{\rm Pl}$, the constraint in Eq.~(\ref{cnstr1}) starts depending mildly on the number of e-folds: for example, taking $f = M_{\rm Pl}$ in Eq.~(\ref{constraint_G}) gives $\alpha \gg 20 N_e^{0.28}$.

We now discuss the condition $T <\Lambda$, that allows us not to consider temperature effects in the axion mass in Eq.~(\ref{axion_mass_T}). Using the equations in Table I for $T$ and $\Lambda$ we obtain
\begin{equation} \label{condition_T_Lambda}
\alpha > \frac{15}{\pi^2}\,\frac{\sin^2\phi_i/f}{\sqrt{1+\cos\phi_i/f}},
\end{equation}
which is always satisfied for all values of $\alpha$, as can be checked numerically. Modeling $\Gamma$ through Eq.~(\ref{decay_constant}) gives $\Gamma_\phi = 0$ and the slow-roll parameter $\beta = 0$ in this model. Summing up, the temperature of the plasma in the NWI model is constrained by $H < T < \Lambda$, with the additional constraint $\Gamma \gg 3H$ that yields to Eq.~(\ref{constraint_G}).

\section{Perturbations from inflation} \label{Perturbations from inflation}

As mentioned in the Introduction, one attractive feature of inflation is that scalar and tensor perturbations emerge during this epoch. These features later evolve into primordial fluctuations of the density profile and gravitational waves, that might leave an imprint in the CMBR anisotropy and on the large scale structures \cite{mukhanov, guth_pi, hawking, starobinsky1, bardeen1}. Each fluctuation is characterized by a power spectrum and a spectral index, respectively $\Delta^2_{\mathcal{R}}(k)$, $n_s$ for density perturbations and $\Delta^2_{\mathcal{T}}(k)$, $n_T$ for tensor perturbations. Here we use results from the warm inflation literature to set constraints on the NWI model.

\subsection{Scalar power spectrum} \label{Scalar power spectrum}

The spectrum of the adiabatic density perturbations generated by inflation is specified by the power spectrum $\Delta^2_{\mathcal{R}}(k)$, which depends mildly on the co-moving wavenumber $k$ according to a spectral index $n_s(k)$ as \cite{kosowsky}
\begin{equation} \label{curvature_perturbations}
\Delta^2_{\mathcal{R}}(k) \equiv \frac{k^3\,P_{\mathcal{R}}(k)}{2\pi^2} = \Delta_{\mathcal{R}}^2(k_0)\,\left(\frac{k}{k_0}\right)^{n_s(k)-1}.
\end{equation}
The function $\Delta_{\mathcal{R}}^2(k)$ describes the contribution to the total variance of primordial curvature perturbations $\mathcal{R}$ due to $\mathcal{R}$ at a given scale per logarithmic interval in $k$ \cite{komatsu}: the WMAP collaboration reports the combined measurement from WMAP+BAO+SN of $\Delta_{\mathcal{R}}^2(k_0)$ at the reference wavenumber $k=k_0 = 0.002 {\rm ~Mpc^{-1}}$, 
\begin{equation} \label{constraint_power_spectrum}
\Delta^2_{\mathcal{R}}(k_0) = (2.445 \pm 0.096) \times 10^{-9},
\end{equation}
where the uncertainty refers to a 68$\%$ likelihood interval. The RHS of Eq.~(\ref{curvature_perturbations}) is evaluated when a given co-moving wavelength crosses outside the Hubble radius during inflation, and the LHS when the same wavelength re-enters the horizon. In Eq.~(\ref{curvature_perturbations}) we have used the notation in Ref.~\cite{komatsu} for the density perturbations. Other authors use the symbol $P_\mathcal{R}(k)$ for our $\Delta^2_\mathcal{R}(k)$ and $\mathcal{R}_k^2$ for our $P_\mathcal{R}(k)$, and might differ for factors of $2\pi^2$.

In both warm and cool inflation models, the scalar power spectrum has the form
\begin{equation}\label{scalar_spectrum}
\Delta^2_{\mathcal{R}}(k) = \left(\frac{H}{\dot{\phi}}\,\langle\delta \phi\rangle\right)^2,
\end{equation}
where $\dot{\phi} \simeq -U_\phi/(3H+\Gamma)$ and $\langle\delta \phi\rangle$ describes the spectrum of fluctuations in the inflaton field. The LHS of Eq.~(\ref{scalar_spectrum}) is computed at the time at which the largest density perturbations on observable scales are produced, corresponding $N_e$ e-foldings before the end of inflation. Quantum fluctuations predict \cite{guth_pi}
\begin{equation}
\langle\delta \phi \rangle_{\rm quantum} = \frac{H}{2\pi},
\end{equation}
while thermal fluctuations provide (we used Eqs.~(21) and~(23) in Ref.~\cite{bastero_gil})
\begin{equation} \label{thermal_fluctuations}
\langle\delta \phi \rangle_{\rm thermal} = \left(\frac{\Gamma\,H\,T^2}{(4\pi)^3}\right)^{1/4}.
\end{equation}
Using Eqs.~(\ref{scalar_spectrum}) and~(\ref{thermal_fluctuations}) for the variance of fluctuations we obtain the scalar power spectrum in the strongly dissipative regime of warm inflation as
\begin{equation} \label{power_spectrum_1}
\Delta^2_{\mathcal{R},{\rm warm}}(k_0) = \frac{1}{(4\pi)^{3/2}}\,\frac{H^{5/2}\,\Gamma^{1/2}T}{\dot{\phi}^2},
\end{equation}
which is the same result as in Refs.~\cite{taylor_berera, moss} with an extra factor $1/(2\pi)^2$ included to account for our normalization of the power spectrum. For comparison, the power spectrum in the usual cool inflation is found as
\begin{equation} \label{power_spectrum_cool}
\Delta^2_{\mathcal{R},{\rm cool}}(k_0) = \left(\frac{H^2}{2\pi\dot{\phi}}\right)^2,
\end{equation}
so that density perturbations are larger in warm inflation by a factor
\begin{equation}\label{def_xi}
\xi \equiv \frac{\Delta^2_{\mathcal{R}}(k_0)}{\Delta^2_{\mathcal{R},{\rm cool}}(k_0)} = \left(\frac{\pi}{4}\,\frac{\Gamma\,T^2}{H^3}\right)^{1/4} \approx 10^9\,\left(\frac{\Gamma_{12}}{m_{\phi\,9}^3\,f_{16}^5}\right)^{1/4}.
\end{equation}

From now on, scalar perturbations are considered only in the warm inflation scenario, so we suppress the index ``warm'' in Eq.~(\ref{power_spectrum_1}). Using the expressions for $H$ and $T$ in Table I we write Eq.~(\ref{power_spectrum_1}) as
\begin{equation} \label{power_spectrum1}
\Delta^2_{\mathcal{R}}(k_0) \simeq \left(\frac{50}{(3\pi)^7}\right)^{1/8}\,g_*^{-1/4}(T)\left(\frac{f\,\Gamma^3}{m_\phi\,M_{\rm Pl}^3}\right)^{3/4}\,\left(\frac{(1+\cos\phi_i/f)^{3/4}}{\sin\phi_i/f}\right)^{3/2}.
\end{equation}
In deriving this last expression we have used the fact that the largest density perturbations are produced when $\phi = \phi_i$ \cite{natural_inflation}. Another form of Eq.~(\ref{power_spectrum1}) that is more useful for computations is
\begin{equation} \label{power_spectrum2}
\Delta^2_{\mathcal{R}}(k_0) \simeq 3.7\times 10^{-13}\,\left(\frac{g_*(T)}{228.75}\right)^{-1/4}\,\alpha^{3/4}\,\Gamma_{12}^{3/2}\,\left(\frac{(1+\cos\phi_i/f)^{3/4}}{\sin\phi_i/f}\right)^{3/2}.
\end{equation}
Eq.~(\ref{power_spectrum2}) defines the power spectrum in terms of $\alpha$, $\Gamma$ and $N_e$, the latter appearing implicitly in the definition of $\phi_i$. 

\begin{figure}[h!]
\begin{center}
\includegraphics[width=10cm]{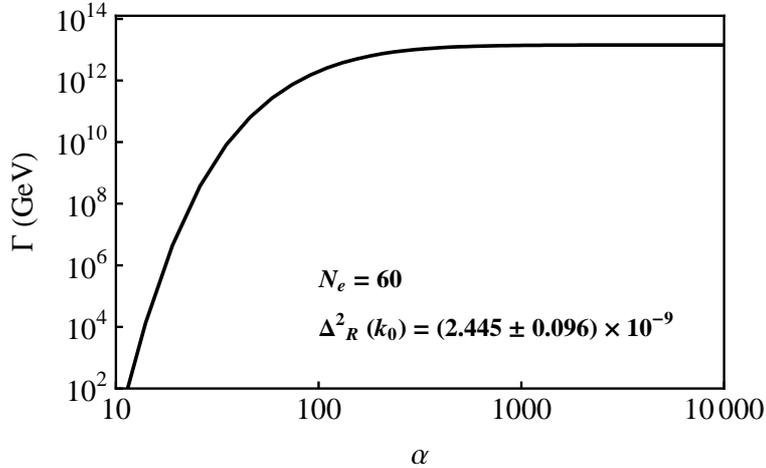}
\caption{The dissipation term $\Gamma$ as a function of $\alpha$, given by Eq.~(\ref{constraint_power_spectrum}) with $\Delta^2_{\mathcal{R}}(k_0) = (2.445 \pm 0.096) \times 10^{-9}$. We used Eq.~(\ref{power_spectrum2}) for the analytic expression of the scalar power spectrum.}
\label{plotGamma}
\end{center}
\end{figure}

We equate the expression for the power spectrum in Eq.~(\ref{power_spectrum2}) to the measured value from WMAP in Eq.~(\ref{constraint_power_spectrum}) to obtain a relation between $\Gamma$ and $\alpha$, see Figure~\ref{plotGamma}. In Figure~\ref{constraint_power_spectrum}, we only consider the case $N_e=60$, while other values of $N_e$ do not modify the curves sensibly unless $\alpha \gg 1$. The thickness of the line approximately corresponds to the one-sigma uncertainty in Eq.~(\ref{constraint_power_spectrum}). From Figure~\ref{plotGamma}, the dissipation $\Gamma$ reaches a maximum value $\Gamma_{\rm max}$ at large values of $\alpha$, which we find analytically from equating Eqs.~(\ref{constraint_power_spectrum}) and~(\ref{power_spectrum2}) in the limit $\alpha \gg 1$,
\begin{equation}\label{gamma_max}
\Gamma_{\rm max} = \frac{3.52\pm 0.09}{\sqrt{N_e}}\,\times 10^{13}{\rm ~GeV}\,\left(\frac{g_*(T)}{228.75}\right)^{1/6}.
\end{equation}

\begin{figure}[h!]
\begin{center}
\includegraphics[width=10cm]{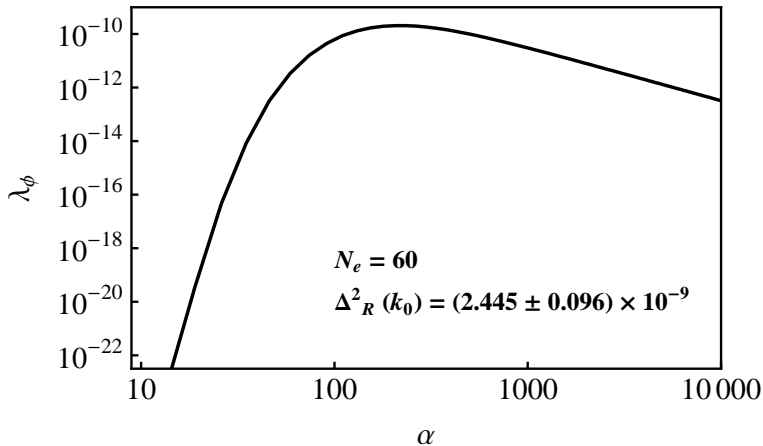}
\caption{The inflaton quartic self-coupling $\lambda_\phi$ as a function of the parameter $\alpha$, Eq.~(\ref{power_spectrum3}). The parameters $\Delta^2_{\mathcal{R}}(k_0)$ and $N_e$ are fixed as indicated in the figure.}
\label{plotLambda}
\end{center}
\end{figure}

To study the strength of the quartic self-interaction in the NWI model, we use Eqs.~(\ref{self_interaction}) and~(\ref{def_alpha}) to eliminate $\Gamma$ in Eq.~(\ref{power_spectrum2}) and obtain a relation between $\lambda_\phi$ and $\alpha$,
\begin{equation} \label{power_spectrum3}
\Delta^2_{\mathcal{R}}(k_0) \simeq 6.89\times 10^{-14}\,\left(\frac{g_*(T)}{228.75}\right)^{-1/4}\,\alpha^{9/4}\,\lambda_\phi^{3/4}\,\left(\frac{(1+\cos\phi_i/f)^{3/4}}{\sin\phi_i/f}\right)^{3/2}.
\end{equation}
The dependence of the quartic self-interaction $\lambda_\phi$ on $\alpha$ thus obtained is shown in Figure~\ref{plotLambda}. We see from this figure that the self-interaction is $\lambda_\phi \lesssim 10^{-10}$ for any $\alpha$, and the NWI model can easily satisfy the constraint for the quartic self-interaction term  $\lambda_\phi \lesssim 10^{-8}$ obtained in Ref.~\cite{limits_self_coupling}.

\subsection{Scalar spectral index}

The scalar spectral index $n_s$ describes the mild dependence of the scalar power spectrum on the wavenumber $k$, as in Eq.~(\ref{curvature_perturbations}). We expand the spectral index around the reference scale $k_0$ as
\begin{equation}
n_s(k) = n_s + \frac{1}{2}\,\tau\,\ln \frac{k}{k_0},
\end{equation}
where $n_s \equiv n_s(k_0)$ and the spectral tilt $\tau$ is
\begin{equation}
\tau = \frac{dn_s(k)}{d\ln k/k_0}\bigg|_{k=k_0}.
\end{equation}
Using Eq.~(\ref{curvature_perturbations}), the scalar spectral index is \cite{kosowsky}
\begin{equation}\label{derivative_power_spectrum}
n_s-1 = \frac{\partial}{\partial \ln k/k_0}\,\ln\frac{\Delta^2_{\mathcal{R}}(k)}{\Delta^2_{\mathcal{R}}(k_0)}.
\end{equation}
In warm inflation, the spectral index is (\cite{hall}; see also Ref.~\cite{berera_spectral_index})
\begin{equation}
n_s -1 = \frac{1}{Q}\left(-\frac{9}{4}\epsilon+\frac{3}{2}\eta-\frac{9}{4}\beta\right),
\end{equation}
while the spectral tilt depends on higher orders in the slow-roll parameters, and for this reason it will be neglected here. With the expressions in Eq.~(\ref{slow_roll_parameters}) for $\epsilon$ and $\eta$, and with $\beta = 0$, we find
\begin{equation}
n_s -1 = \frac{3}{16\pi\,G\,U^2\,Q}\left(U_{\phi\phi}\,U-\frac{3}{4}U_\phi^2\right),
\end{equation}
or, using $U(\phi)$ in Eq.~(\ref{potential}) and its derivatives at $\phi = \phi_i$, together with the values of $Q$, $\epsilon$ and $\eta$ in Table 1,
\begin{equation} \label{ns_eq}
n_s  = 1 -\frac{3}{8}\sqrt{\frac{3}{2\pi}}\,\frac{m_\phi\,M_{\rm Pl}}{\Gamma\,f}\frac{3+\cos\phi_i/f}{\sqrt{1+\cos\phi_i/f}} = 1 -\frac{1.50}{\alpha}\frac{3+\cos\phi_i/f}{\sqrt{1+\cos\phi_i/f}}.
\end{equation}
This expression differs from what found in the ``cool'' Natural Inflation scenario (see Eq.~(11) in Ref.~\cite{natural_inflation2}).

\begin{figure}[h!]
\begin{center}
  \includegraphics[width=10cm]{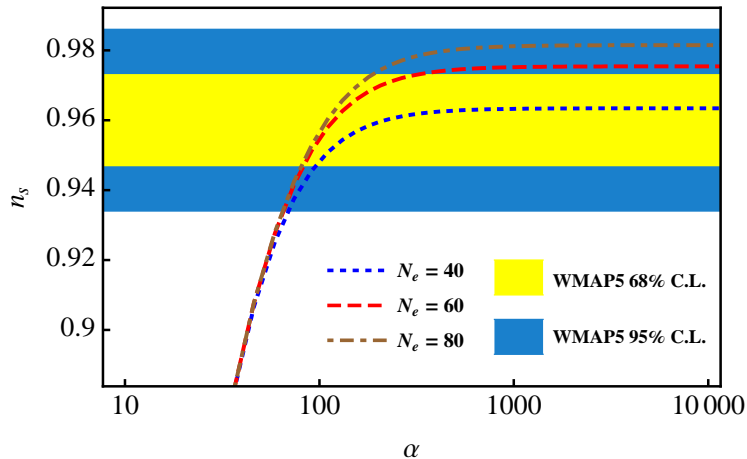}
\caption{The scalar spectral index $n_s$, Eq.~(\ref{ns_eq}), for different values of the number of e-folds. Blue dotted: $N_e=40$; Red dashed: $N_e=60$; Brown dot-dashed: $N_e=80$. Also shown are the C.L. regions from the combined WMAP+BAO+SN measurement, Eq.~(\ref{ns_exp}). Yellow: 68\%; violet: 95\%.}
\label{plot_ns}
\end{center}
\end{figure}

Figure~\ref{plot_ns} shows the dependence of the spectral index on $\alpha$ as in Eq.~(\ref{ns_eq}), for different values of the number of e-folds $N_e$. We have compared the values of $n_s$ with the  combined WMAP +SN+BAO measurement \cite{komatsu},
\begin{equation} \label{ns_exp}
n_s = 0.960 \pm 0.013 \quad\hbox{at 68\% C.L.},
\end{equation}
so that the shaded yellow region in Figure~\ref{plot_ns} corresponds to the 68\% C.L., while the light blue region is the corresponding 95\% C.L. region. We see that small values of $N_e \approx 50$ and large values of $\alpha > O(100)$ are favored, although values of $N_e = 60-70$ can be accommodated in the 68\% C.L. region for $100\lesssim \alpha\lesssim 300$ or in the 95\%C.L. region for all values of $\alpha$.

For large values of $\alpha$ the scalar spectral index $n_s$ is essentially independent of $\alpha$, while for $\alpha \lesssim N_e$ it is the dependence on $N_e$ that vanishes. This behavior can be summed up by considering Eq.~(\ref{ns_eq}) in these two limits,
\begin{equation} \label{ns_approx}
1 - n_s = \begin{cases}
4.2/\alpha & \hbox{$\alpha \lesssim N_e$},\\
1.5/N_e & \hbox{$\alpha\gtrsim N_e$}.
\end{cases}
\end{equation}
This type of dependence of $n_s$ on either $N_e$ or $\alpha$ has been also noticed in models of Natural Inflation set in the usual cool inflation scenario, see for example Eq.~(12) in Ref.~\cite{natural_inflation2}. In the case in Ref.~\cite{natural_inflation2}, in which there is no dissipation term $\Gamma$, and the inflaton mass $m_\phi$ is adjusted so that $\Lambda\approx 10^{16}$ GeV, the result analogous to Eq.~(\ref{ns_approx}) is expressed in terms of $f$ instead of $\alpha$.

\subsection{Tensor power spectrum}

Since warm inflation considers thermal fluctuations instead of quantum fluctuations to generate scalar perturbations, it is only density fluctuations that modify in this scenario while tensor perturbations show the same spectrum as in the usual cool inflation \cite{moss}. Defining the tensor power spectrum as
\begin{equation} \label{tensor_perturbations}
\Delta^2_{\mathcal{T}}(k) \equiv \frac{k^3\,P_{\mathcal{T}}(k)}{2\pi^2} = \Delta_{\mathcal{T}}^2(k_0)\,\left(\frac{k}{k_0}\right)^{n_T},
\end{equation}
where the tensor spectral index $n_T$ is assumed to be independent of $k$, because current measurement cannot constraint its scale dependence.
WMAP does not constraint $\Delta_{\mathcal{T}}^2(k_0)$ directly, but rather the tensor-to-scalar ratio
\begin{equation}
r \equiv \frac{\Delta_{\mathcal{T}}^2(k_0)}{\Delta_{\mathcal{R}}^2(k_0)}.
\end{equation}
which qualitatively measures the amplitude of gravitational waves per density fluctuations. The WMAP+BAO+SN measurement constraints the tensor-to-scalar ratio as \cite{komatsu}
\begin{equation}
r < 0.22 \quad \hbox{at 95$\%$ C.L.}.
\end{equation}

Since in warm inflation the scalar power spectrum is enhanced by the quantity $\xi$ in Eq.~(\ref{def_xi}) with respect to the value in cool inflation, the tensor-to-scalar ratio and thus gravitational waves are reduced in the NWI model by the same amount $\xi$. Using Eq.~(\ref{power_spectrum_1}) for the scalar power spectrum and the expression for the tensor power spectrum,
\begin{equation}
\Delta_{\mathcal{T}}^2(k_0) = \frac{16\,H^2}{\pi M_{\rm pl}^2},
\end{equation}
the tensor-to-scalar ratio in warm inflation is
\begin{equation}\label{r_measure}
r = \frac{128\sqrt{\pi}}{M_{\rm Pl}^2}\,\frac{\dot{\phi}^2}{\sqrt{H\Gamma}\,T}.
\end{equation}
With the values in Table 1 we find
$$r \simeq 128\,\left(\frac{\pi^7}{150}\right)^{1/8}\,g_*^{1/4}(T)\,\left(\frac{f^5\,m_{\phi}^{11}}{\Gamma^9\,M_{\rm Pl}^7}\right)^{1/4}\,\frac{(\sin\phi_i/f)^{3/2}}{(1+\cos\phi_i/f)^{1/8}} = $$
\begin{equation}
= 1.62 \times 10^{-13}\,\left(\frac{g_*(T)}{228.75}\right)^{1/4}\,\left(\frac{f_{16}^5\,m_{\phi\,9}^{11}}{\Gamma_{12}^9}\right)^{1/4}\,\frac{(\sin\phi_i/f)^{3/2}}{(1+\cos\phi_i/f)^{1/8}}.
\end{equation}
In the limit $\alpha \gg N_e$, with $\phi_i \approx \pi-\sqrt{8}N_e/\alpha$, we obtain
\begin{equation}
r = 4.63 \times 10^{-14}\,N_e^{5/4}\left(\frac{g_*(T)}{228.75}\right)^{1/4}\,\left(\frac{m_{\phi\,9}}{\Gamma_{{\rm max},12}}\right)^{4},\quad\hbox{for $\alpha \gg N_e$}.
\end{equation}
Using the WMAP measure of the tensor-to-scalar ratio in Eq.~(\ref{r_measure}), we constraint the inflaton mass in the NWI model to
\begin{equation} \label{bound_r}
m_\phi < 2.7\times 10^{12}{\rm ~GeV}\,\left(\frac{r}{0.22}\right)^{1/4}\,\left(\frac{N_e}{60}\right)^{-13/16}\,\left(\frac{g_*(T)}{228.75}\right)^{5/48},
\end{equation}
where we used the expression for $\Gamma_{\rm max}$ in Eq.~(\ref{gamma_max}). Future measurements will not substantially improve the bound in Eq.~(\ref{bound_r}), because of the power $1/4$ raising $r$. As an example, the forecast PLANCK measurement will constraint the tensor-to-scalar ratio by one order of magnitude with respect to WMAP, $r \lesssim 0.01$ \cite{planck}, thus the bound in Eq.~(\ref{bound_r}) will approximately lower by a factor of two.

\begin{figure}[h!]
\begin{center}
\includegraphics[width=10cm]{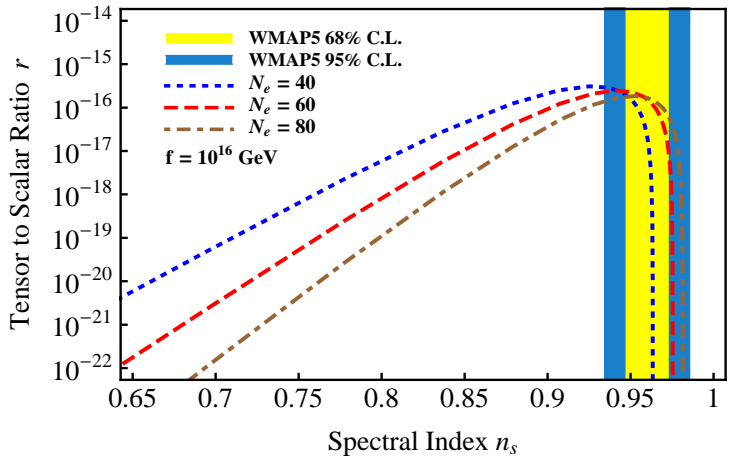}
\caption{Prediction from the NWI model and WMAP constraints in the $r$-$n_s$ plane for $f=10^{16}$ GeV. Blue dotted line: $N_e=40$; Red dashed line: $N_e=60$; Brown dot-dashed line: $N_e=80$. The yellow and violet regions are the parameter spaces allowed by WMAP+BAO+SN at 68\% and 95\% C.L. respectively.}
\label{plot16}
\end{center}
\end{figure}

\begin{figure}[h!]
\begin{center}
\includegraphics[width=10cm]{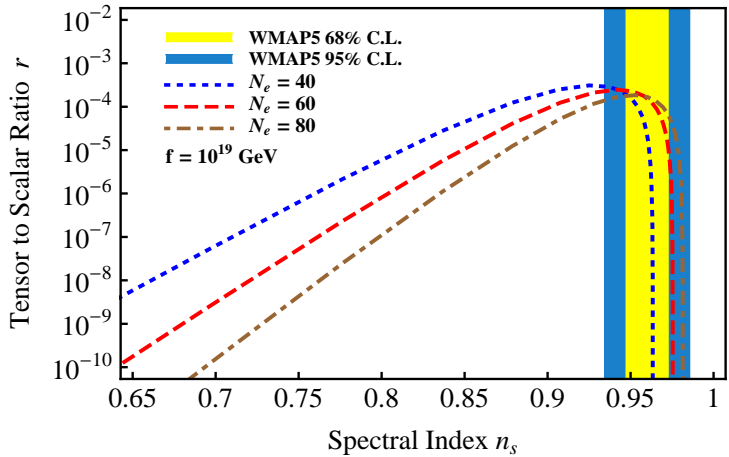}
\caption{Prediction from the NWI model and WMAP constraints in the $r$-$n_s$ plane for $f~=~10^{19}{\rm~GeV}$. Blue dotted line: $N_e=40$; Red dashed line: $N_e=60$; Brown dot-dashed line: $N_e=80$. The yellow and violet regions are the parameter spaces allowed by WMAP+BAO+SN at 68\% and 95\% C.L. respectively.}
\label{plot19}
\end{center}
\end{figure}

Results are summarized in Figures~\ref{plot16} and~\ref{plot19}. In Figure~\ref{plot16} we show the constraints on the NWI model from the WMAP data in the $r$-$n_s$ plane, in the case $f = 10^{16}{\rm~GeV}$. The WMAP data are better in agreement with a low number of e-folds $N_e = 40-50$, although values up to $N_e = 80$ can be accommodated within the 95\% C.L. region. We conclude that a value of the axion decay constant of the order of the GUT scale $f \sim \Lambda_{\rm GUT}$ can be easily embedded in the NWI model, making it possible to construct microscopic theories of warm inflation with axion-like particles at the GUT scale. For this value of the axion decay constant, the expected value of $r$ and thus the amount of gravitational waves that are produced with this type of inflation is extremely low. If gravitational waves are found with $r \sim 10^{-14}$ or above, this model has to be abandoned and a $\Lambda_{\rm GUT}$ valued axion decay constant $f$ is no longer viable.

For comparison, we show the constraints on the NWI model from the WMAP data in the $r$-$n_s$ plane when $f = 10^{19}{\rm~GeV}$ in Figure~\ref{plot19}. We see that the expected amount of gravitational waves has raised with respect to the case in Figure~\ref{plot16} by ten orders of magnitude, while data still favor relatively low values of the number of e-folds. For a given value of $N_e$, the curves in Figures~\ref{plot16} and~\ref{plot19} rigidly move along the vertical $r$ axis when changing the parameter $f_a$.

\section{Discussion and conclusion}\label{discussion}

The parameter space of the inflaton field in the NWI model is broader than that in the usual Natural Inflation, because of the presence of the extra quantity $\Gamma$ describing dissipation and of the enlarged slow-roll conditions in Eq.~(\ref{slow_roll}).

The NWI model allows to lower the value of the axion decay constant $f$ from the Planck scale resulting in Natural Inflation to the GUT scale; a ratio $f/M_{\rm Pl} \approx 10^{-3}$ helps overcoming some difficulties encountered in Natural Inflation model-building.

A decay constant of the order of the Planck scale is still possible, since $f$ is not bound from above. In the case $f \sim M_{\rm Pl}$, Natural Inflation and NWI can be distinguished observationally by a measurement of the tensor-to-scalar ratio $r$, see Figures~\ref{plot16} and~\ref{plot19}. This difference in the value of the ratio $r$ comes from the fact that in the NWI model the amplitude of gravitational waves is suppressed by the factor $\xi$ in Eq~(\ref{def_xi}) with respect to the standard cool inflation.

Measurements of the scalar spectral index $n_s$ favor larger values of $\alpha \gtrsim N_e$, as shown in Figure~\ref{plot_ns}. In this region, the value of $\phi_i$ in Eq.~(\ref{definition_phi1}) can be approximated with $\phi_i \approx \pi - \sqrt{8}N_e/\alpha$, see Eq.~(\ref{phi1_approx}), and the axion-like potential in Eq.~(\ref{potential}) cannot be distinguished from a pure quadratic one through the dynamic of the inflaton. In this condition, one would need to obtain the value of the self-interaction $\lambda_\phi$ independently, for example from considering density fluctuations in the CMBR; Natural Inflation and NWI models predict a precise ratio between the height of the potential and the self-interaction coupling constant, $U/\lambda_\phi = 12\,f^2$, whereas other inflaton models predict different values of this ratio.

The possibility that the axion-like energy scale $f$ is of the order of the GUT scale when warm inflation is considered has been already taken into account in Ref.~\cite{mohanty}. However, we differ from Ref.~\cite{mohanty} in various points: for example, we do not consider the temperature as an independent variable, because in warm inflation the radiation bath is thermalized and temperature is expressed is linked to the radiation energy density $\rho_r$ as in Eq.~(\ref{definition_radiation}). Moreover, here we have included a detailed analysis of the CMBR observables $r$, $\Delta^2_{\mathcal{R}}(k)$ and $n_s$ that was missing in Ref.~\cite{mohanty}.

We have presented a model in which Natural Inflation takes place within the warm inflation scenario: in such model, the decay constant $f$ no longer ties to the Planck scale as in the usual Natural Inflation model, but it can be as low as the GUT scale, $f \sim 10^{16}$ GeV. We have shown the viability of the NWI model and its agreement with current astrophysical data.

During the completion of the present paper, the work in Ref.~\cite{mishra} where a similar model is considered came to our attention. The authors in Ref.~\cite{mishra} constrained the Natural Warm Inflation model using data from the WMAP experiment and obtained values of $f$ that are generally below the Planck scale, in agreement with the results in this paper.

\begin{acknowledgments}
The author would like to thank Paolo Gondolo for useful discussions and comments on this paper, and Akhilesh Nautiyal for finding a mistake in an earlier version of this paper.
\end{acknowledgments}

\end{document}